# Coupled feedback loops maintain synaptic long-term potentiation: A computational model of PKMzeta synthesis and AMPA receptor trafficking

Peter Helfer[1*] and Thomas R. Shultz[2]

[1] Department of Psychology and Integrated Program in Neuroscience, McGill University, Montreal, Quebec, Canada

[2] Department of Psychology and School of Computer Science, McGill University, Montreal, Quebec, Canada

[*] Corresponding author

E-mail: peter.helfer@mail.mcgill.ca

# Abstract

In long-term potentiation (LTP), one of the most studied types of neural plasticity, synaptic strength is persistently increased in response to stimulation. Although a number of different proteins have been implicated in the sub-cellular molecular processes underlying induction and maintenance of LTP, the precise mechanisms remain unknown. A particular challenge is to demonstrate that a proposed molecular mechanism can provide the level of stability needed to maintain memories for months or longer, in spite of the fact that many of the participating molecules have much shorter life spans. Here we present a computational model that combines simulations of several biochemical reactions that have been suggested in the LTP literature and show that the resulting system does exhibit the required stability. At the core of the model are two interlinked feedback loops of molecular reactions, one involving the atypical protein kinase PKMζ and its messenger RNA, the other involving PKMζ and GluA2-containing AMPA receptors. We demonstrate that robust bistability – stable equilibria both in the synapse's potentiated and unpotentiated states – can arise from a set of simple molecular reactions. The model is able to account for a wide range of empirical results, including induction and maintenance of late-phase LTP, cellular memory reconsolidation and the effects of different pharmaceutical interventions.

# Author summary

The brain stores memories by adjusting the strengths of connections between neurons, a phenomenon known as synaptic plasticity. Different types of plasticity mechanisms have either a strengthening or a weakening effect and produce synaptic modifications that last from milliseconds to months or more. One of the most studied forms of plasticity, long-term potentiation, is a persistent increase of synaptic strength that results from stimulation and is believed to play an important role in both short-term and long-term memory. Researchers have identified many proteins and other molecules involved in long-term potentiation and formulated different hypotheses about the biochemical processes underlying its induction and maintenance. A growing number of studies support an important role for the protein PKMζ (protein kinase M Zeta) in long-term potentiation. To investigate the explanatory power of this hypothesis, we built a computational model of the proposed biochemical reactions that involve this protein and ran simulations of a number of experiments that have been reported in the literature. We find that our model is able to explain a wide range of empirical results and thus provide insights into the molecular mechanisms of memory.

# Introduction

The brain stores memories by adjusting the strengths of connections between neurons. Such synaptic plasticity comes in different forms that strengthen or weaken synapses and range from very short-lived to long-lasting. One of the most well-studied forms of plasticity is long-term potentiation, LTP, a phenomenon whereby synaptic strength is persistently increased in response to stimulation. Different forms of LTP are known to play important roles in both short-term and long-term memory.

Many different proteins have been identified in the sub-cellular molecular processes that are involved in LTP. An important question is how these proteins, with lifetimes measured in hours or days, can maintain memories for months or years. We present a computational model that demonstrates how this problem can be solved by two interconnected feedback loops of molecular reactions.

We begin with an overview of LTP with emphasis on the empirical findings that our model aims to explain. This is followed by a description of the model, an account of our results, and discussion of their implications.

## Background

In his address to the Royal Society in 1894, Santiago Ramon y Cajal hypothesized that the brain stores information by adjusting the strengths of associations between neurons, as well as by growing new connections [1]. In the years since, the existence of both of these mechanisms, now known as synaptic plasticity and synaptogenesis, respectively, has been well established, and there is ample evidence that synaptic plasticity plays an important role in learning and memory [2–4].

Neurons communicate by transmitting signals across chemical synapses, where presynaptic axon terminals connect to postsynaptic neurons, most often on their dendrites. When a nerve impulse (action potential) arrives at the axon terminal, neurotransmitter molecules are released into the synaptic cleft, a narrow gap between the two neurons, where they activate receptors in the membrane of the postsynaptic neuron. This sets in motion a series of biochemical events in the postsynaptic neuron, the details of which depend on the type of receptor, among other factors. Synaptic strength depends both on the amount of transmitter that is released by the arrival of a nerve impulse at the axon terminal and on the number and sensitivity of the receptors. It may thus be regulated on either the pre- or postsynaptic side, and mechanisms of synaptic plasticity have been shown to operate in both compartments [3]. Plasticity may either strengthen or weaken a synapse, and the effect may be short-lived or long-lasting. Short-term synaptic plasticity, lasting from milliseconds to minutes, is primarily due to presynaptic mechanisms that adjust the amount of transmitter release, whereas postsynaptic modifications that adjust the number and sensitivity of receptors are important for long-term plasticity [4]. In particular, this is true of long-term potentiation (LTP), a type of persistent strengthening of synapses in response to stimulation [5,6], which has been studied extensively in the CA3-CA1 synapses of the rodent hippocampus [4] and is known to depend on an increase in the number of receptors inserted in the postsynaptic membrane [7].

There are at least two forms of LTP: Moderately strong stimulation induces early-phase LTP (E-LTP), which persists for at most a few hours. When the stimulation is stronger, E-LTP may be followed by late-phase LTP (L-LTP), which can last for days, months or longer [7,8] and is believed to be an important mechanism for the storage of long-term memories [9,10]. The establishment of L-LTP, known as synaptic or cellular memory consolidation, is a process that takes less than an hour [11,12] and requires synthesis of new protein. This has been demonstrated by showing that infusion of protein-synthesis-inhibiting drugs such as anisomycin can prevent establishment of L-LTP [12–15]. On the behavioral level, protein synthesis inhibition (PSI) has been shown to impair the formation of long-term memory, consistent with the notion of L-LTP as a memory mechanism [16].

Once long-term memory is established, it is in general no longer vulnerable to infusion of a protein synthesis inhibitor [16]. However, memory retrieval can induce a state of transient instability, during which the memory is again susceptible to protein synthesis inhibition [17–19]. This susceptibility of memory to post-retrieval PSI infusion has been shown to correlate with instability of L-LTP at the neural level [20,21], providing further evidence of the importance of LTP as a mechanism of long-term memory. The synaptic destabilization that is triggered by memory retrieval is followed by a period of restabilization which has similarities with the initial synaptic consolidation that follows memory acquisition. It has therefore become known as *memory reconsolidation* [19], more specifically *synaptic* (or *cellular*) *reconsolidation*, to avoid confusion with the related but distinct phenomenon *systems reconsolidation*, a temporary dependence on the hippocampus for restabilization of a memory after reactivation (retrieval). For reviews of reconsolidation research, see [22–24]. For a computational model of systems consolidation and reconsolidation, see [25].

## Glutamatergic synapses

In this report, we focus on L-LTP induction and maintenance at glutamatergic synapses, the most abundant type of synapse in the vertebrate nervous system [26,27]. Glutamatergic synapses contain several kinds of receptors that are activated by the neurotransmitter glutamate. Of particular interest for LTP are the α-amino-3-hydroxy-5-methyl-4-isoxazolepropionic acid receptor (AMPA receptor or AMPAR), which mediates synaptic transmission [28], and the N-methyl-D-aspartate receptor (NMDA receptor or NMDAR), which is involved with regulatory functions including the regulation of synaptic strength [29,30].

AMPARs are ion channels that open when activated by the neurotransmitter glutamate. The opening of the channel allows positively charged ions, mainly sodium and potassium, to flow through the cell membrane [31]. This causes a partial depolarization of the membrane, which at rest is polarized by a net negative charge inside the cell. The partial depolarization is known as an excitatory postsynaptic potential, or EPSP, and the amplitude of the EPSP produced by a single action potential arriving at a synapse is a measure of synaptic strength. Among other factors, the EPSP amplitude depends on the number of AMPARs inserted in the postsynaptic density (PSD), the area of cell membrane that constitutes the receiving side of the synapse [31]. Thus mechanisms that control the trafficking of AMPARs into and out of the PSD play an important part in the regulation of synaptic strength.

AMPARs are heterotetramers, i.e. they consist of four non-identical subunits. The subunits are of four different kinds, named GluA1, GluA2, GluA3 and GluA4, and AMPARs can be made up of different combinations of these [32]. GluA2 is of particular interest here, because L-LTP is associated with an increase in the number of GluA2-containing AMPARs inserted in the PSD [20,33,34].

AMPA receptors are not permanently inserted in the PSD, but are constantly being recycled. Certain proteins transport AMPARs into the PSD from pools maintained in adjacent areas, while others remove them (a process known as internalization or endocytosis) and either recycle them to stand-by pools or mark them for degradation [35,36].

## Protein kinase M zeta (PKMζ)

Many proteins have been implicated in the induction and maintenance of LTP, including CaMKII, PKA, MAPK and several isoforms of PKC (for a review, see [7]). An atypical isoform of PKC, protein kinase Mζ (PKMζ), is believed to play an important role for L-LTP. The level of PKMζ has been shown to increase as the result of NMDA receptor stimulation [37,38], consistent with its proposed role in L-LTP induction. Inhibition of PKMζ activity results in disruption of established L-LTP [39–41], and perfusion of PKMζ into a neuron can induce L-LTP [39] . PKMζ activity is believed to increase the number of inserted GluA2-containing AMPARs at the synapse both by facilitating the trafficking of these receptors into the PSD and by inhibiting their removal [42]. GluA2-containing AMPARs are held at extrasynaptic pools by the protein PICK1 which binds to the GluA2 subunit [34]. PKMζ facilitates interaction between the trafficking protein NSF and the GluA2 subunit, which results in its release from PICK1, freeing the AMPARs to diffuse laterally into the PSD [34]. Furthermore, once GluA2-containing AMPARs are inserted in the PSD membrane, PKMζ prevents their removal by inhibiting the interaction between the protein BRAG2 and the GluA2 subunit [43], an interaction that plays a key part in endocytosis of GluA2-containing AMPARs [42,44].

While GluA2-containing AMPARs are important for the stabilization of L-LTP, there is evidence that GluA2-lacking AMPARs play an important role in the induction of early-phase LTP (E-LTP), and also in reconsolidation. Several studies have shown that GluA2-lacking AMPARs are initially inserted at the time of memory acquisition or LTP induction, and then gradually replaced by GluA2-containing AMPARs during consolidation [45–47]. Hong et al. [20] showed that memory reactivation triggers an abrupt replacement of GluA2-containing AMPARs by GluA2-lacking AMPARs. This is followed by a gradual reversal, i.e. the GluA2-containing AMPARs are restored and the number of GluA2-lacking AMPARs declines, as the potentiated state of the synapse is restabilized [20]. Because the temporary removal of GluA2-containing AMPARs is compensated for by an increase in GluA2-lacking AMPARs, the synaptic strength remains more or less constant during the period of instability [20]. Rao-Ruiz et al. [21] reported similar results, although they observed a brief period of reduced synaptic strength between the GluA2-containing AMPAR removal and GluA2-lacking AMPAR insertion. Taken together, these results suggest that the stabilization of LTP, both initially during consolidation, and after reactivation-induced destabilization, requires insertion of GluA2-containing AMPARs, and that PKMζ plays an important role in maintaining the GluA2-containing AMPARs at the synapse.

An important question is how L-LTP, which can last for months or longer [8], can be maintained by a protein like PKMζ, with a half-life that probably does not exceed several hours or at most a few days [48–51]. A proposed answer to this question involves local translation of messenger RNA (mRNA) in or near dendritic spines. Most synapses are formed at dendritic spine heads, with one synapse per spine [52]. It has been shown that PKMζ mRNA is transported from the cell body to dendrites [53,54], but the mRNA in its basal state is translationally repressed by molecules that bind to it, or to the complex of proteins required to initiate translation [50,53,55]. There is evidence that PKMζ catalyzes reactions that lift this translational block [49,56], possibly through inhibition of the PIN1 protein [42], resulting in a positive feedback loop [49]. By promoting its own synthesis in this manner, PKMζ may be able to remain at an increased level, and thus maintain L-LTP, for a long time, perhaps indefinitely.

It has also been suggested that the increased amount of inserted GluA2-containing AMPARs at a potentiated synapse captures the PKMζ molecules and keeps them from dissipating away from the synaptic compartment [42]. This hypothesis is supported by several studies that show that blocking endocytosis of GluA2-containing AMPARs can prevent depotentiation under protocols that otherwise cause disruption of L-LTP [21,33,57]. Together with PKMζ's inhibiting effect on AMPAR endocytosis this constitutes a second feedback loop, a reciprocal relationship in which PKMζ and GluA2-containing AMPARs prevent each other's removal from the synapse. As we shall see, the interaction between these two feedback loops plays a central role in our explanation of synaptic bistability, that is that synapses have two stable equilibrium states, unpotentiated and potentiated. Transient stimuli can cause a synapse to transition between these two states, but in the absence of such signals it tends to remain in one state or the other.

## L-LTP, LTM and pharmacological interventions

The notion that L-LTP is an important neural correlate of long-term memory (LTM) has been supported experimentally by demonstrating that pharmacological interventions that block L-LTP induction also interfere with the establishment of LTM [58], and that interventions that disrupt established L-LTP also impair consolidated memories [59]. Here we consider three types of pharmaceuticals that have been shown to produce significant results with respect to both L-LTP induction and maintenance, and to related behavior-level memory phenomena.

**Protein synthesis inhibitors.** Infusion of protein synthesis inhibitors (PSIs) such as anisomycin into brain tissue can prevent the induction of L-LTP [58], and also interferes with memory consolidation, the establishment of LTM [60,61]. Once L-LTP is established, it becomes resistant to infusion of anisomycin [11,12]. This does not mean that L-LTP can be maintained indefinitely without ongoing protein synthesis, but rather that it can tolerate an interruption of protein synthesis for the amount of time that anisomycin remains active after infusion.

Reactivation of a consolidated memory, e.g. by a reminder, can temporarily return it to a labile state in which it is again vulnerable to PSI infusion [18,60]. The putative molecular process underlying this phenomenon has been termed *cellular* or *synaptic memory reconsolidation* [18,62]. Concordant with the hypothesis that L-LTP is the neural correlate of LTM, the temporary post-reactivation vulnerability of LTM to PSI infusion can be explained as destabilization of L-LTP, followed by a restabilization phase that requires protein synthesis, hence the susceptibility to PSI. The destabilization has been shown to require the activity of NMDA receptors [29], and to depend critically on endocytosis of GluA2-containing AMPARs [57,63].

Thus protein synthesis inhibition is known to both prevent establishment of L-LTP and to block reconsolidation, i.e. block restabilization of L-LTP after retrieval-induced destabilization.

**ZIP.** Much of the work demonstrating the role of PKMζ in L-LTP is based on administration of the synthetic peptide ZIP (zeta-inhibitory peptide), which binds to the catalytic region of the PKMζ molecule, thus blocking its enzymatic activity [41]. On the behavioral level, infusion of ZIP into brain tissue has been shown to impair consolidated LTM [59]. On the neural level, ZIP is known to disrupt established L-LTP when applied during the maintenance phase [39–41,64]. These results are consistent with the notion of a positive feedback loop: Inhibiting PKMζ's enzymatic activity prevents it from catalyzing its own synthesis; the PKMζ concentration then drops, the AMPAR endocytosis rate increases, and the synapse returns to its basal state. On the other hand, ZIP does not prevent L-LTP induction when applied only during or immediately after stimulation. This was demonstrated by Ren et al. [65] in an in-vitro experiment where onset and duration of ZIP application were precisely controlled.

**GluA2$_{3Y}$.** GluA2$_{3Y}$ is a synthetic peptide that blocks regulated endocytosis of GluA2-containing AMPARs [66,67]. Infusion of GluA2$_{3Y}$ has been shown to block both the destabilizing effect of PSI infusion after memory reactivation [20,57] and the depotentiating effect of ZIP during L-LTP maintenance [33]. The GluA2$_{3Y}$ peptide is modeled on a sequence of the GluA2 subunit's carboxyl tail and its endocytosis-inhibiting effect is believed to be due to competitive disruption of the binding of endocytosis-related proteins to this sequence on GluA2 subunits [68].

## Computational model

The findings described above suggest a model of L-LTP maintenance with two connected feedback loops: (1) PKMζ maintains its own mRNA in a translatable state and translation of the mRNA in turn replenishes PKMζ. (2) PKMζ maintains GluA2-containing AMPARs at the synapse, and these in turn keep PKMζ molecules from dissipating away from the synaptic compartment. Below we describe a computational model that incorporates these relationships and investigate its ability to account for results reported in the empirical literature.

# Methods

## Deterministic vs. stochastic simulation

Systems of chemical reactions can be modeled either by deterministic methods based on ordinary differential equations (ODEs) or by stochastic simulation. When the numbers of molecules are small, stochastic simulation is the better choice, because random fluctuations then have significant effects that are not captured by deterministic methods [69]. In particular, random fluctuations can cause a small system to spontaneously transition from one steady state to another; the resulting impact on system stability can be studied in a stochastic simulation, but not in a deterministic model [70], because the latter only accounts for average reaction rates over a large number of molecules.

The molecules of interest for our simulation are present in small numbers in a dendritic spine head, e.g. fewer than a hundred PKMζ molecules (see S1 Text) and at most ca 150 AMPARs [71,72]. This is well below the size of system that can be realistically simulated by deterministic methods [70,73]. We therefore base our simulation on the Gillespie algorithm [74], a well-established and widely used approach to discrete and stochastic simulation of reaction systems [69,70,73].

## Model description

The model consists of four inter-dependent pairs of processes (see Fig 1):

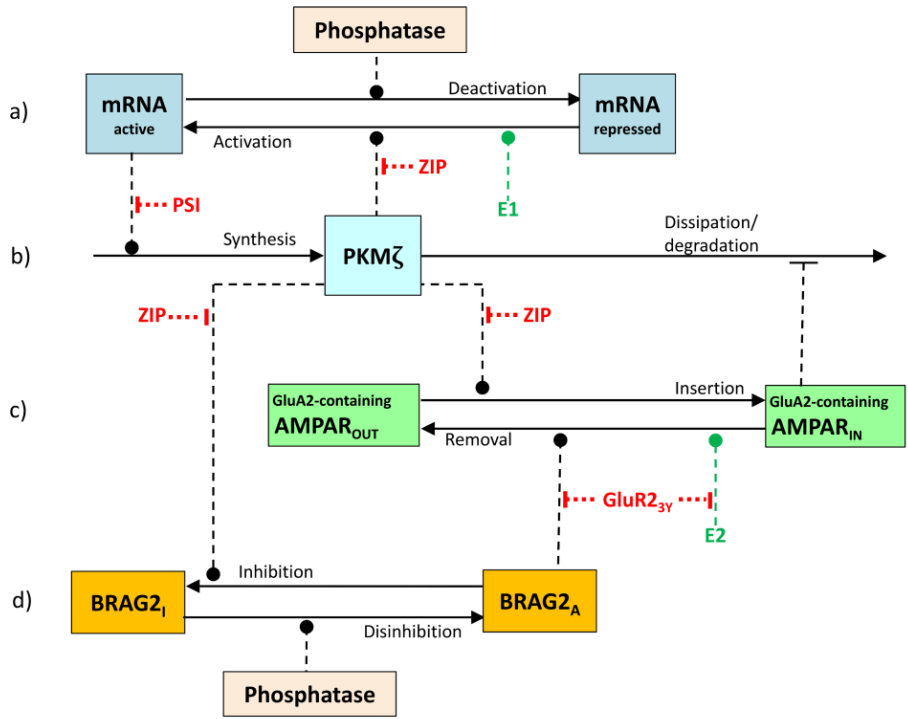

**Fig 1: Process diagram.** a) Activation/deactivation of PKMζ mRNA (blue).
Translational repression of mRNA is lifted by catalytic activity of PKMζ, possibly by
phosphorylation of mRNA-binding proteins. A phosphatase (pink) dephosphorylates the
same proteins, returning mRNA to its repressed state. b) Synthesis and
degradation/dissipation of PKMζ (cyan). Synthesis consists in local translation of PKMζ
mRNA. Degradation and/or dissipation away from the synaptic compartment is inhibited
by inserted GluA2-containing AMPARs. c) Trafficking of GluA2-containing AMPARs
(green) into and out of the PSD. Insertion is facilitated by PKMζ, and removal
(endocytosis) by the BRAG2 protein. d) Inhibition/disinhibition of BRAG2-GluA2
interaction. Inhibition is modeled as phosphorylation of BRAG2 (orange) catalyzed by
PKMζ, and disinhibition as dephosphorylation catalyzed by a phosphatase. E1 and E2
(dark green) are enzymes activated by NMDAR stimulation at L-LTP induction and
memory reactivation, respectively. The effects of PSI, ZIP and GluA2$_{3Y}$ (red) are modeled by disabling the indicated catalytic reactions. Solid arrows represent chemical reactions and receptor trafficking. Dashed lines with filled circles represent catalytic activity. Dashed lines with crossbars represent inhibition.

**Activation/deactivation of PKMζ mRNA.** PKMζ lifts the constitutive translational repression of PKMζ mRNA by phosphorylating some substrate, possibly mRNA-binding proteins attached to the mRNA. The mRNA molecule with attached proteins and ribosomes (polysome) is represented as a single molecule in the model, and de-repression is modeled as phosphorylation of (some component of) this molecule by PKMζ. The opposite reaction, dephosphorylation by a phosphatase assumed to be present at fixed concentration, returns the mRNA to its repressed state.

**Synthesis and degradation/dissipation of PKMζ.** Synthesis consists in local translation of PKMζ mRNA. (This is somewhat speculative: PKMζ mRNA has been shown to be present in dendrites [53,54], but not specifically in dendritic spines.) Inserted GluA2-containing AMPARs inhibit degradation and/or dissipation of PKMζ away from the synaptic compartment by binding PKMζ molecules [42], probably via a scaffold protein such as PICK1 or KIBRA [75]. This is modeled as an affinity of PKMζ for inserted GluA2-containing AMPARs, with a reduced dissipation/degradation rate while so attached.

**GluA2-containing AMPAR trafficking into and out of the PSD.** The model includes a fixed-size population of GluA2-containing AMPARs. At any time, a subset of the AMPARs are inserted in the PSD while the remainder are maintained in extrasynaptic pools. Transport of AMPARs into the PSD is facilitated by PKMζ and removal (endocytosis) is enabled by the protein BRAG2. In addition to these two regulated processes, constitutive processes traffic AMPARs into and out of the synapse at lower rates.

**Inhibition and disinhibition of BRAG2-GluA2 interaction.** The mechanism by which PKMζ inhibits the interaction between BRAG2 and the GluA2 subunit to block AMPAR removal from the PSD is not known, but presumably involves phosphorylation of some substrate. We model the inhibition as phosphorylation of the BRAG2 molecule itself; other possibilities include phosphorylation of a site on the GluA2 subunit or of another participating protein. The BRAG2-GluA2 interaction is restored through dephosphorylation of the same substrate by a phosphatase, which is assumed to be present in fixed concentration.

Although the increase in PKMζ level that is associated with L-LTP induction is known to depend on NMDAR activation [38], the underlying biochemical pathways are unknown. In the model this mechanism is represented by an unspecified enzyme that we call E1 which, when activated by a reaction cascade triggered by NMDAR activation, has the ability to lift the translational block on PKMζ mRNA, thereby enabling PKMζ synthesis.

Similarly, the destabilizing effect of memory reactivation has been shown to depend on NMDAR activity and on endocytosis of GluA2-containing AMPARs [20,57,76], but the biochemical cascades that connect these event have not yet been identified. In our model, reactivation is simulated as an increase in the level of a second unspecified enzyme E2 with the ability to catalyze endocytosis of GluA2-containing AMPAR.

In addition to these processes, the model includes simulation of the effects of the three pharmaceuticals described in the introduction. The time intervals that these drugs remain at a high enough concentration to inhibit their targets depend on the doses infused and also on their specific rates of decay or metabolism. The intervals used here are based on activity periods reported in the cited references:

**PSI:** Infusion of a protein synthesis inhibitor is simulated by disabling PKMζ synthesis for nine hours, the amount of time that the protein synthesis inhibitor anisomycin remains active after infusion into brain tissue [77].

**ZIP:** Administration of the ZIP peptide is simulated by disabling PKMζ's enzymatic activity – catalysis of mRNA activation, facilitation of GluA2-containing AMPAR trafficking into the PSD and inhibition of BRAG2-GluA2 interaction – for twelve hours [78].

**GluA2$_{3Y}$:** Perfusion of GluA2$_{3Y}$ is simulated by disabling regulated endocytosis of GluA2-containing AMPAR for twelve hours [76]. (GluA2$_{3Y}$ does not affect constitutive endocytosis of GluA2-containing AMPAR [67].)

Table 1 lists the molecule species included in the model, including complexes formed during enzymatic reactions. All simulations begin in the lower (unpotentiated) steady state with the indicated initial molecule counts.

**Table 1: Molecule species**

| Symbol | Description | Initial count |
|--------|-------------|---------------|
| P | Unbound PKMζ | 0 |
| $R_I$ | unphosphorylated PKMζ mRNA (inactive) | 100 |

| $R_A$ | phosphorylated PKMζ mRNA (active) | 0 |
|---|---|---|
| PP | phosphatase | 100 |
| $PP \cdot R_A$ | PP + $R_A$ complex | 0 |
| $E1_A$ | E1 enzyme, active | 0 |
| $E1_I$ | E1 enzyme, inactive | 100 |
| $E1_A \cdot R_I$ | $E1_A$ + $R_I$ complex | 0 |
| $A_U$ | Uninserted GluA2-containing AMPAR | 100 |
| $A_I$ | Inserted GluA2-containing AMPAR | 0 |
| $A_I \cdot P$ | PKMζ bound to inserted AMPAR | 0 |
| $P \cdot R_I$ | P + $R_I$ complex | 0 |
| $A_I \cdot P \cdot R_I$ | $A_I$ + P + $R_I$ complex | 0 |
| $B_A$ | Active BRAG2 | 100 |
| $B_I$ | Inactive BRAG2 | 0 |
| $PP \cdot B_I$ | PP + $B_I$ complex | 0 |
| $P \cdot B_A$ | P + $B_A$ complex | 0 |
| $A_I \cdot P \cdot B_A$ | $A_I$ + P + $B_A$ complex | 0 |
| $B_A \cdot A_I$ | $B_A$ + $A_I$ complex | 0 |
| $B_A \cdot A_I \cdot P$ | $B_A$ + $A_I$ + P complex | 0 |

| E2$_A$ | E2 enzyme, active | 0 |
|---|---|---|
| E2$_I$ | E2 enzyme, inactive | 100 |

## Simulated Reactions

**Activation of PKMζ mRNA.** PKMζ mRNA is present in dendritic spines, but is translationally repressed in its basal state [42,53] due to mRNA-binding proteins that prevent translation from being initiated [55]. PKMζ is able to lift the repression, possibly by phosphorylating these proteins, thus catalyzing its own synthesis in a positive feedback loop. We model mRNA with its associated proteins as a single molecule, represented by $R_I$ in its inactive repressed state, and by $R_A$ when activated. Activation is modeled using Michaelis-Menten kinetics [73], i.e. a PKMζ molecule ($P$) and an inactive mRNA molecule ($R_I$) form a complex $P \cdot R_I$. The complex may then either dissociate (reaction 2) or the catalytic reaction (3) may take place, producing active mRNA ($R_A$):

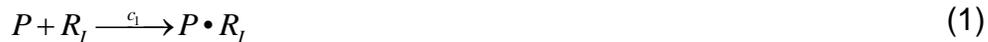
$$P + R_I \xrightarrow{c_1} P \cdot R_I \tag{1}$$

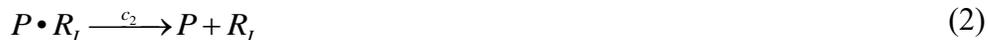
$$P \cdot R_I \xrightarrow{c_2} P + R_I \tag{2}$$

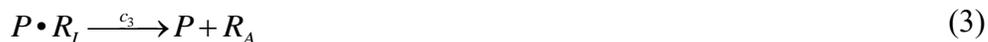
$$P \cdot R_I \xrightarrow{c_3} P + R_A \tag{3}$$

**Deactivation of PKMζ mRNA.** The PKMζ mRNA returns to its repressed state when the mRNA-binding proteins are dephosphorylated by a phosphatase which we denote by PP. This is also modeled with Michaelis-Menten kinetics (as are all enzymatic reactions in the model):

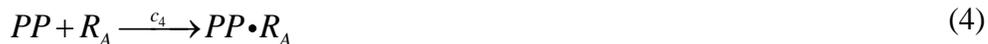
$$PP + R_A \xrightarrow{c_4} PP \cdot R_A \tag{4}$$

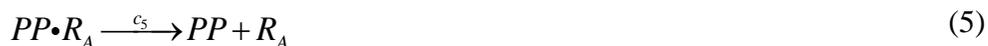
$$PP \cdot R_A \xrightarrow{c_5} PP + R_A \tag{5}$$

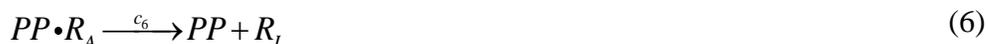
$$PP \cdot R_A \xrightarrow{c_6} PP + R_I \tag{6}$$

**PKMζ synthesis and degradation/dissipation.** PKMζ is synthesized by local translation of active mRNA (reaction 7). Over time PKMζ degrades or diffuses away from the synaptic compartment. Reaction 8 represents the combined effect of these two processes. The model is unspecific with respect to their relative importance for PKMζ turnover.

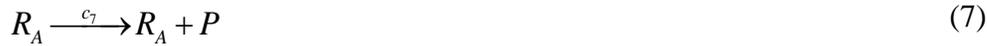
$$R_A \xrightarrow{\ c_7\ } R_A + P \tag{7}$$

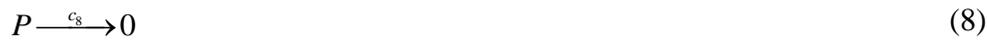
$$P \xrightarrow{\ c_8\ } 0 \tag{8}$$

**Inhibition/disinhibition of BRAG2.** BRAG2 is inhibited by PKMζ and reactivated by phosphatase. Both processes are described by Michaelis-Menten kinetics. $B_A$ and $B_I$ denote active and inhibited BRAG2, respectively:

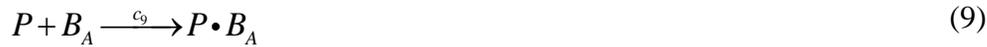
$$P + B_A \xrightarrow{\ c_9\ } P \bullet B_A \tag{9}$$

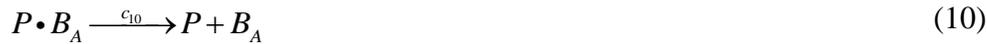
$$P \bullet B_A \xrightarrow{\ c_{10}\ } P + B_A \tag{10}$$

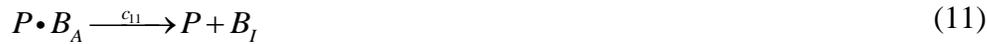
$$P \bullet B_A \xrightarrow{\ c_{11}\ } P + B_I \tag{11}$$

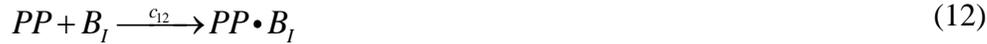
$$PP + B_I \xrightarrow{\ c_{12}\ } PP \bullet B_I \tag{12}$$

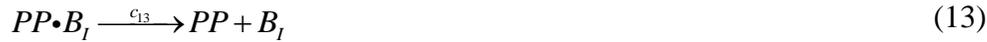
$$PP \bullet B_I \xrightarrow{\ c_{13}\ } PP + B_I \tag{13}$$

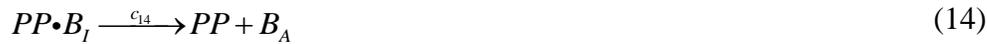
$$PP \bullet B_I \xrightarrow{\ c_{14}\ } PP + B_A \tag{14}$$

**AMPA receptor trafficking.** Transport of GluA2-containing AMPARs into the PSD has been shown to involve a trafficking process that is facilitated by PKMζ [34]. Because the details of this process are unknown, including which substrate of PKMζ mediates it, we model it as a simple enzymatic reaction wherein PKMζ catalyzes the conversion of an uninserted GluA2-containing AMPAR, $A_U$, to an inserted one, $A_I$.

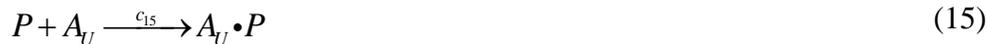
$$P + A_U \xrightarrow{\ c_{15}\ } A_U \bullet P \tag{15}$$

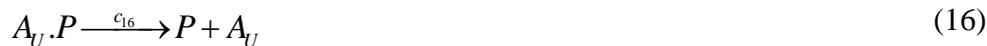
$$A_U . P \xrightarrow{\ c_{16}\ } P + A_U \tag{16}$$

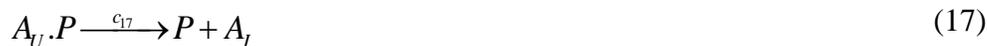
$$A_U . P \xrightarrow{\ c_{17}\ } P + A_I \tag{17}$$

The protein BRAG2 catalyzes endocytosis of GluA2-containing AMPARs, removal from the PSD.

$$B_A + A_I \xrightarrow{c_{18}} B_A \bullet A_I \tag{18}$$

$$B_A \bullet A_I \xrightarrow{c_{19}} B_A + A_I \tag{19}$$

$$B_A \bullet A_I \xrightarrow{c_{20}} B_A + A_U \tag{20}$$

A pair of unregulated processes maintain background cycling of GluA2-containing AMPARs into and out of the PSD:

$$A_U \xrightarrow{c_{21}} A_I \tag{21}$$

$$A_I \xrightarrow{c_{22}} A_U \tag{22}$$

**Sequestering of PKMζ in the synaptic compartment.** Our model implements the notion suggested by Sacktor [42] and supported by empirical results [20,33,57], that GluA2-containing AMPARs, when inserted in the PSD, prevent diffusion of PKMζ molecules away from the synapse and/or slows down degradation of PKMζ. We model this as a PKMζ molecule binding to an inserted GluA2-containing AMPAR to form a complex $A_I \bullet P$ (reaction 23) and by bound PKMζ having a much lower rate of dissipation/degradation than free PKMζ ($c_{24} << c_8$):

$$P + A_I \xrightarrow{c_{23}} A_I \bullet P \tag{23}$$

$$A_I \bullet P \xrightarrow{c_{24}} A_I \tag{24}$$

The $A_I \bullet P$ complex is dissolved if the GluA2-containing AMPAR is removed from the membrane by BRAG2 or constitutively:

$$B_A + A_I \bullet P \xrightarrow{c_{25}} B_A \bullet A_I \bullet P \tag{25}$$

$$B_A \bullet A_I \bullet P \xrightarrow{c_{26}} B_A + A_I \bullet P \tag{26}$$

$$B_A \bullet A_I \bullet P \xrightarrow{c_{27}} B_A + A_U + P \tag{27}$$

$$A_I \bullet P \xrightarrow{c_{28}} A_u + P \tag{28}$$

PKMζ remains catalytically active while sequestered by GluA2-containing AMPARs, thus the reactions catalyzed by free PKMζ (reactions 1-3 and 9-11) are also catalyzed by PKMζ when it is bound to $A_I$:

$$A_I \bullet P + R_I \xrightarrow{c_{29}} A_I \bullet P \bullet R_I \tag{29}$$

$$A_I \bullet P \bullet R_I \xrightarrow{c_{30}} A_I \bullet P + R_I \tag{30}$$

$$A_I \bullet P \bullet R_I \xrightarrow{c_{31}} A_I \bullet P + R_A \tag{31}$$

$$A_I \bullet P + B_A \xrightarrow{c_{32}} A_I \bullet P \bullet B_A \tag{32}$$

$$A_I \bullet P \bullet B_A \xrightarrow{c_{33}} A_I \bullet P + B_A \tag{33}$$

$$A_I \bullet P \bullet B_A \xrightarrow{c_{34}} A_I \bullet P + B_I \tag{34}$$

**NMDAR stimulation.** The mechanism by which NMDAR activation causes an increase in PKMζ is unknown. We model the effect of strong NMDAR stimulation as a rapid increase in the number of active molecules of an unspecified enzyme E1 which, like PKMζ, activates PKMζ mRNA. $E1_I$ and $E1_A$ represent the E1 enzyme in its active and inactive states, respectively:

$$E1_A + R_I \xrightarrow{c_{35}} E1_A \bullet R_I \tag{35}$$

$$E1_A \bullet R_I \xrightarrow{c_{36}} E1_A + R_I \tag{36}$$

$$E1_A \bullet R_I \xrightarrow{c_{37}} E1_A + R_A \tag{37}$$

The E1 enzyme spontaneously deactivates at a rate that is specified by the reaction constant $c_{38}$:

$$E1_A \xrightarrow{c_{38}} E1_I \tag{38}$$

**Reactivation.** Reactivation of a consolidated memory causes it to become destabilized [18,79,80]. The molecular mechanism underlying this destabilization is not well understood, but has been showed to depend critically on endocytosis of GluA2-containing AMPAR [20,57,63]. We model the destabilizing effect of reactivation as an increase in the number of molecules of a second unspecified enzyme, E2, which catalyzes AMPAR endocytosis:

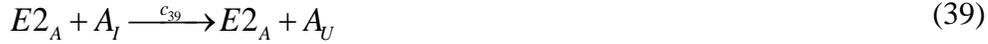

$$E2_A + A_I \xrightarrow{c_{39}} E2_A + A_U \tag{39}$$

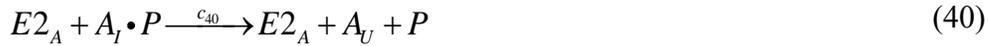

$$E2_A + A_I \bullet P \xrightarrow{c_{40}} E2_A + A_U + P \tag{40}$$

As in the case of BRAG2-catalyzed endocytosis (reaction 27), the AMPAR/PKMζ complex dissolves when the AMPAR is endocytosed (reaction 40).

The $E_2$ enzyme spontaneously deactivates at a rate that is specified by the reaction constant $c_{41}$:

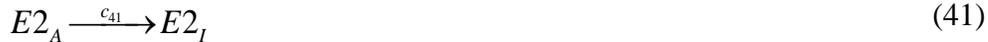

$$E2_A \xrightarrow{c_{41}} E2_I \tag{41}$$

**Protein synthesis inhibition.** The effect of PSI infusion is simulated by disabling synthesis of PKMζ (reaction 7).

**Inhibition of PKMζ by ZIP.** The effect of ZIP infusion is simulated by disabling all PKMζ enzymatic activity (reactions 1, 9, 15, 29 and 32).

**Inhibition of AMPAR endocytosis by GluA2$_{3Y}$.** The effect of GluA2$_{3Y}$ infusion is simulated by disabling regulated AMPAR endocytosis, whether catalyzed by BRAG2 (reactions 18 and 25) or by the E2 enzyme (reactions 39 and 40).

The simulated reactions are summarized in Table 2. Reaction rates are controlled by Gillespie reaction constant, $c_1$, $c_2$, etc., such that $c_i\, dt$ is the average probability that a particular combination of the reactant molecules of reaction $i$ will react during the next infinitesimal time interval $dt$ [74]. The values for the reaction constants have been selected so that the model's behavior approximates the observed time courses of the simulated experiments; see cited references in the description of each simulation.

**Table 2: Simulated reactions**

| | Reaction | Description | $c_i$ (s$^{-1}$) |
|---|---|---|---|
| 1 | $P + R_I \xrightarrow{c_1} P \bullet R_I$ | Formation of P•R$_I$ complex | 10.0 |
| 2 | $P \bullet R_I \xrightarrow{c_2} P + R_I$ | Dissolution of P•R$_I$ complex | 400.0 |
| 3 | $P \bullet R_I \xrightarrow{c_3} P + R_A$ | Activation of PKMζ mRNA, catalyzed by PKMζ | 100.0 |
| 4 | $PP + R_A \xrightarrow{c_4} PP \bullet R_A$ | Formation of PP•R$_A$ complex | 4.0 |
| 5 | $PP \bullet R_A \xrightarrow{c_5} PP + R_A$ | Dissolution of PP•R$_A$ complex | 400.0 |
| 6 | $PP \bullet R_A \xrightarrow{c_6} PP + R_I$ | Deactivation of PKMζ mRNA, catalyzed by phosphatase | 100.0 |
| 7 | $R_A \xrightarrow{c_7} R_A + P$ | Translation of PKMζ mRNA | 0.2 |
| 8 | $P \xrightarrow{c_8} 0$ | PKMζ degradation or dissipation | 0.65 |
| 9 | $P + B_A \xrightarrow{c_9} P \bullet B_A$ | Formation of P•B$_A$ complex | 1.0 |
| 10 | $P \bullet B_A \xrightarrow{c_{10}} P + B_A$ | Dissolution of P•B$_A$ complex | 400.0 |
| 11 | $P \bullet B_A \xrightarrow{c_{11}} P + B_I$ | Inhibition of BRAG2, catalyzed by PKMζ | 20.0 |
| 12 | $PP + B_I \xrightarrow{c_{12}} PP \bullet B_I$ | Formation of PP•B$_I$ complex | 1.0 |
| 13 | $PP \bullet B_I \xrightarrow{c_{13}} PP + B_I$ | Dissolution of PP•B$_I$ complex | 400.0 |
| 14 | $PP \bullet B_I \xrightarrow{c_{14}} PP + B_A$ | Disinhibition of BRAG2, catalyzed by phosphatase | 0.06 |
| 15 | $P + A_U \xrightarrow{c_{15}} A_U \bullet P$ | Formation of P•A$_U$ complex | 0.4 |
| 16 | $A_U . P \xrightarrow{c_{16}} P + A_U$ | Dissolution of P•A$_U$ complex | 400.0 |
| 17 | $A_U . P \xrightarrow{c_{17}} P + A_I$ | PKMZ-catalyzed trafficking of GluA2-containing AMPAR into the PSD | 20.0 |
| 18 | $B_A + A_I \xrightarrow{c_{18}} B_A \bullet A_I$ | Formation of B$_A$•A$_I$ complex | 10.0 |
| 19 | $B_A \bullet A_I \xrightarrow{c_{19}} B_A + A_I$ | Dissolution of B$_A$•A$_I$ complex | 400.0 |
| 20 | $B_A \bullet A_I \xrightarrow{c_{20}} B_A + A_U$ | BRAG2-catalyzed endocytosis of GluA2-containing AMPAR | 4.0 |
| 21 | $A_U \xrightarrow{c_{21}} A_I$ | Unregulated trafficking of GluA2-containing AMPAR into the PSD | 0.05 |
| 22 | $A_I \xrightarrow{c_{22}} A_U$ | Unregulated removal GluA2-containing AMPAR from the PSD | 0.005 |
| 23 | $P + A_I \xrightarrow{c_{23}} A_I \bullet P$ | Inserted GluA2-containing AMPAR binds PKMζ | 1.0 |
| 24 | $A_I \bullet P \xrightarrow{c_{24}} A_I$ | Degradation of PKMζ bound to inserted AMPAR | 0.0001 |
| 25 | $B_A + A_I \bullet P \xrightarrow{c_{25}} B_A \bullet A_I \bullet P$ | Formation of B$_A$•A$_I$•P complex | 10.0 |

| 26 | $B_A \bullet A_I \bullet P \xrightarrow{c_{26}} B_A + A_I \bullet P$ | Dissolution of BA•AI•P complex | 400.0 |
|---|---|---|---|
| 27 | $B_A \bullet A_I \bullet P \xrightarrow{c_{27}} B_A + A_U + P$ | BRAG2-catalyzed endocytosis of GluA2-containing AMPAR with bound PKMζ. | 4.0 |
| 28 | $A_I \bullet P \xrightarrow{c_{28}} A_u + P$ | Unregulated endocytosis of GluA2-containing AMPAR with bound PKMζ. | 0.005 |
| 29 | $A_I \bullet P + R_I \xrightarrow{c_{29}} A_I \bullet P \bullet R_I$ | Formation of $A_I \bullet P \bullet R_I$ complex | 10.0 |
| 30 | $A_I \bullet P \bullet R_I \xrightarrow{c_{30}} A_I \bullet P + R_I$ | Dissolution of $A_I \bullet P \bullet R_I$ complex | 400.0 |
| 31 | $A_I \bullet P \bullet R_I \xrightarrow{c_{31}} A_I \bullet P + R_A$ | Activation of PKMζ mRNA, catalyzed by AMPAR-bound PKMζ | 100.0 |
| 32 | $A_I \bullet P + B_A \xrightarrow{c_{32}} A_I \bullet P \bullet B_A$ | Formation of $A_I \bullet P \bullet B_A$ complex | 1.0 |
| 33 | $A_I \bullet P \bullet B_A \xrightarrow{c_{33}} A_I \bullet P + B_A$ | Dissolution of $A_I \bullet P \bullet B_A$ complex | 400.0 |
| 34 | $A_I \bullet P \bullet B_A \xrightarrow{c_{34}} A_I \bullet P + B_I$ | Inhibition of BRAG2, catalyzed by AMPAR-bound PKMζ | 20.0 |
| 35 | $E1_A + R_I \xrightarrow{c_{35}} E1_A \bullet R_I$ | Formation of $E1_A \bullet R_I$ complex | 10.0 |
| 36 | $E1_A \bullet R_I \xrightarrow{c_{36}} E1_A + R_I$ | Dissolution of $E1_A \bullet R_I$ complex | 400.0 |
| 37 | $E1_A \bullet R_I \xrightarrow{c_{37}} E1_A + R_A$ | Activation of PKMζ mRNA, catalyzed by E1 enzyme | 100.0 |
| 38 | $E1_A \xrightarrow{c_{38}} E1_I$ | Spontaneous deactivation of E1 enzyme | 0.3 |
| 39 | $E2_A + A_I \xrightarrow{c_{39}} E2_A + A_U$ | Endocytosis of GluA2-containing AMPAR, catalyzed by E2 enzyme | 0.1 |
| 40 | $E2_A + A_I \bullet P \xrightarrow{c_{40}} E2_A + A_U + P$ | E2-catalyzed endocytosis of GluA2-containing AMPAR with bound PKMζ | 0.1 |
| 41 | $E2_A \xrightarrow{c_{41}} E2_I$ | Spontaneous deactivation of E2 enzyme | 0.5 |

## Simulation environment

The model is implemented as a C++ program and all simulations were executed on an Intel i5-2400 computer running the Debian Linux 8.4 operating system.

## Objectives

Our computational model simulates the regulation of PKMζ concentration at the postsynaptic density and its role in the induction and maintenance of L-LTP. The goal for the model is to simulate the empirical results described in the introduction and summarized in Table 3 below. Most of the cited results are from studies of Schaffer collateral synapses on CA1 pyramidal neurons in the rat or mouse hippocampus, a few refer to unspecified hippocampal regions or amygdala of rat or mouse.

**Table 3: Simulation objectives**

| | Result | Description | Citations |
|---|---|---|---|
| 1 | **Induction by NMDAR stimulation** | Strong NMDAR stimulation induces L-LTP | [40,81] |
| 2 | **PSI blocks NMDAR-triggered L-LTP induction** | Infusion of protein synthesis inhibitors prevents L-LTP induction by NMDAR stimulation | [9,82] |
| 3 | **ZIP during stimulation does not prevent L-LTP induction** | ZIP treatment during and immediately after stimulation does not prevent establishment of L-LTP | [65] |
| 4 | **Induction by PKMζ perfusion** | Perfusion of PKMζ into a neuron induces L-LTP | [39,83] |
| 5 | **PSI does not disrupt established L-LTP/LTM** | Application of a protein synthesis inhibitor during L-LTP maintenance (without preceding reactivation) does not cause disruption of L-LTP | [12,18,79] |

| 6 | **Reactivation does not disrupt LTM** | Memory reactivation does not by itself disrupt LTM | [18,79] |
| 7 | **Reactivation followed by PSI infusion does disrupt LTM** | PSI administered within a time window after reactivation disrupts LTM | [18,79] |
| 8 | **GluA2$_{3Y}$ blocks the LTM-disrupting effect of PSI** | GluA2$_{3Y}$ administered together with PSI after reactivation blocks the LTM-disrupting effect of PSI | [20,57,63] |
| 9 | **ZIP disrupts established L-LTP** | Infusion of ZIP during the maintenance phase disrupts L-LTP | [39–41] |
| 10 | **GluA2$_{3Y}$ blocks the depotentiating effect of ZIP** | GluA2$_{3Y}$ infused together with ZIP prevents depotentiation of established L-LTP | [33] |

# Results

In the following plots of simulation results, P denotes the total number of PKMζ molecules in the synaptic compartment, whether free or bound to a substrate or to an AMPAR (see Table 1), and $A_I$ denotes the number of AMPARs inserted in the PSD, with and without bound PKMζ molecules. Reaction numbers refer to the reactions described in Table 2.

## NMDAR stimulation induces L-LTP

We model the result of strong NMDAR stimulation as a rapid increase of the population of active E1 enzyme molecules. This causes the translational repression of PKMζ mRNA to be lifted (reactions 35-37) and synthesis of PKMζ to start (reaction 7). Fig 2 shows a trace of the number of PKMζ molecules, active PKMζ mRNA molecules and GluA2-containing AMPARs inserted in the PSD during a single simulation run.

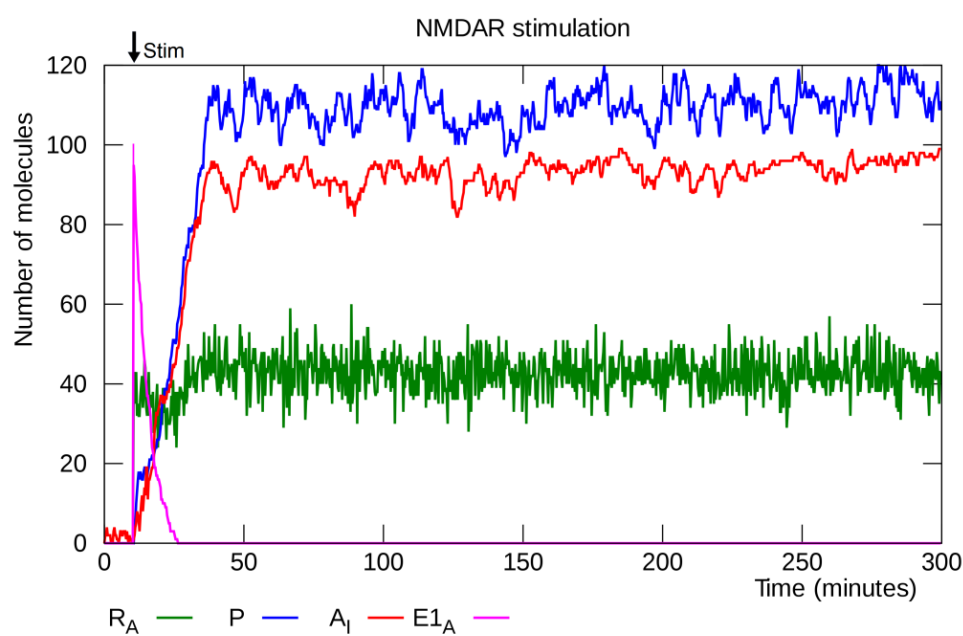

**Fig 2: L-LTP induction, single simulation trace.** NMDAR stimulation is simulated by instantaneous activation of 100 E1 molecules at "Stim". E1 lifts the translational inhibition of PKMζ mRNA, synthesis of PKMζ starts, PKMζ drives up the number of inserted GluA2-containing AMPARs, and the synapse switches to its potentiated steady

state. $R_A$: active PKMζ mRNA, P: PKMζ, $A_I$: inserted GluA2-containing AMPARs, $E1_A$: activated E1 enzyme.

The model has two stable states: an unpotentiated state in which there are very few active mRNA molecules, PKMζ molecules and inserted GluA2-containing AMPARs, and a potentiated state with significantly higher levels of each of these molecules. The brief spike of E1 enzyme lifts the translational repression of enough PKMζ mRNA molecules to trigger a transition to the potentiated state. Although the molecule numbers fluctuate in the potentiated state, it is in fact very stable: No spontaneous depotentiation events are observed even when the model is allowed to run for a full year of simulated time. Fig 3 shows mean molecule counts for 100 simulations of L-LTP induction. It takes the model between 30 and 60 minutes of simulated time to complete the switch to its upper (potentiated) steady state in which there is a high number of inserted GluA2-containing AMPARs. This is consistent with the observed duration of the cellular consolidation window [16,58].

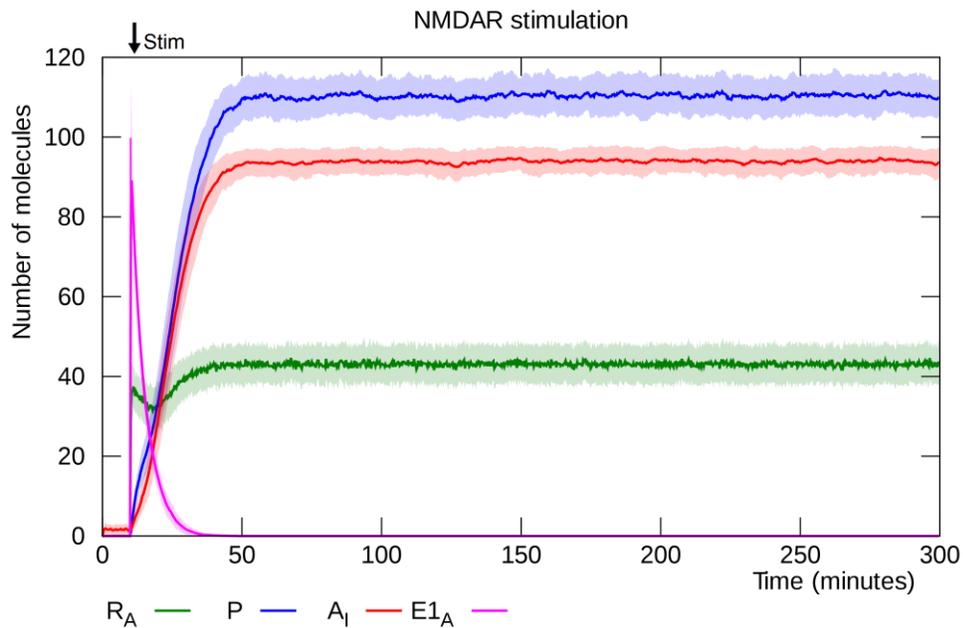

**Fig 3: L-LTP induction.** The same simulation as in Fig 2, but here solid lines represent mean molecule counts for 100 simulations. Lightly colored bands indicate standard deviation. NMDA stimulation triggers a brief spike of E1 activity that activates PKMζ mRNA. This is followed by a slight decline in the number of active mRNA molecules, until the growing amount of PKMζ drives it back up and an equilibrium is reached. $R_A$: active PKMζ mRNA, P: PKMζ, $A_I$: inserted GluA2-containing AMPARs, $E1_A$: activated E1 enzyme.

## PSI prevents L-LTP induction

Simulated PSI infusion prevents NMDAR stimulation from inducing L-LTP, (Fig 4).

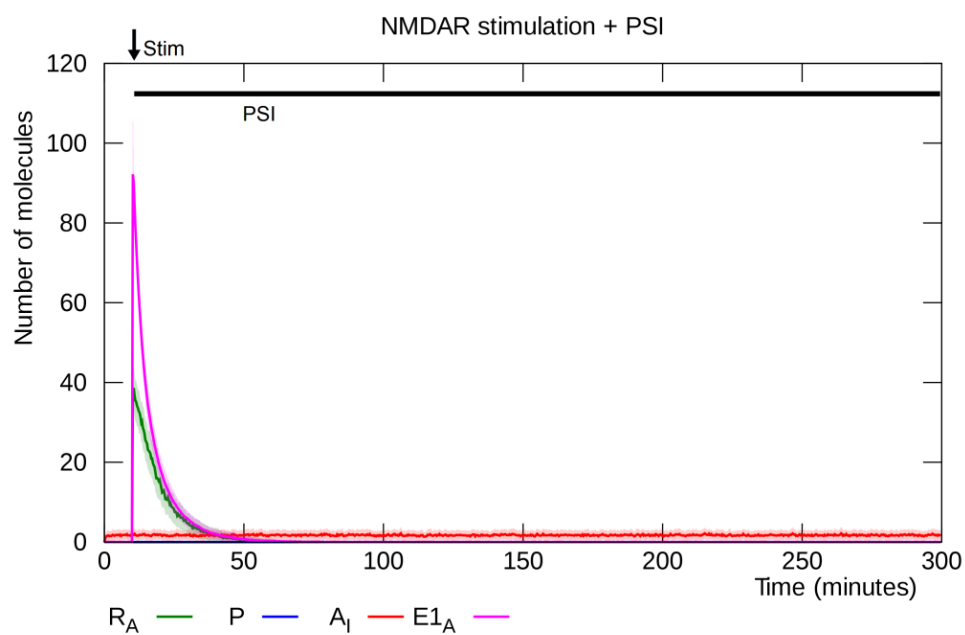

**Fig 4: PSI prevents NMDAR stimulation from inducing L-LTP.** E1$_A$ enzyme activates PKMζ mRNA, but PSI prevents PKMζ synthesis and when the E1 enzyme becomes inactivate, phosphatase returns the mRNA to its inhibited state. Solid lines represent mean molecule counts for 100 simulations. Lightly colored bands indicate standard deviation. R$_A$: active PKMζ mRNA, P: PKMζ, A$_I$: inserted GluA2-containing AMPARs, E1$_A$: activated E1 enzyme.

Although the spike of activated E1 enzyme releases the translational block of mRNA, resulting in a high level of activated PKMζ mRNA (R$_A$ in the model), translation is prevented by the protein synthesis inhibitor, and PKMζ synthesis is not initiated [9,37]. When the E1 enzyme returns to its inactive form the mRNA becomes repressed again, and the model remains in its unpotentiated state. Like the potentiated state, the unpotentiated state is very stable: No spontaneous potentiation events are observed even when running the model for a year of simulated time.

By introducing a variable delay between stimulation and PSI infusion, we can study the model's consolidation window, the time interval after induction during which PSI prevents establishment of L-LTP. As shown in Fig 5, when the delay before PSI infusion is 20 minutes or less, the model consistently settles in the lower (unpotentiated) steady state with zero or very few inserted GluA2-containing AMPARs. When the delay is 50 minutes or more, the model settles in the upper (potentiated) state where the number of inserted GluA2-containing AMPARs fluctuates between ca 60 and 100 (cf. Fig 2). With intermediate delays, the probability of settling in the upper state gradually increases with increasing delay. The model's consolidation window is thus in the range 30 to 45 minutes, consistent with empirical results [11,12]. Fig 5 illustrates the model's bistable character: It settles either in the unpotentiated or potentiated state, never in the region with intermediate numbers of inserted GluA2-containing AMPARs. See also Fig 3 and Fig 4.

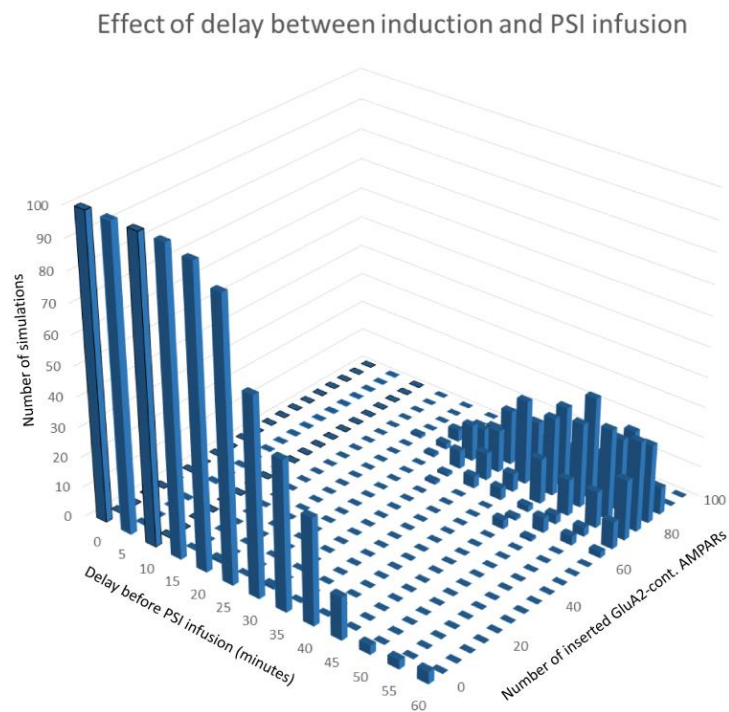

Effect of delay between induction and PSI infusion

**Fig 5: Consolidation window.** Results of simulated NMDAR stimulation followed by PSI infusion after a delay varying from 0 to 60 minutes in 5-minute steps. One hundred simulations were run with each value for the delay. The number of inserted GluA2-containing AMPARs was recorded twenty hours after stimulation. For each value of the delay, the heights of the columns indicate the number of simulations that terminated with the corresponding numbers of inserted GluA2-containing AMPARs.

## ZIP during and immediately after stimulation does not prevent L-LTP induction

ZIP application during stimulation and the first 10 minutes thereafter after does not prevent L-LTP induction, (Fig 6).

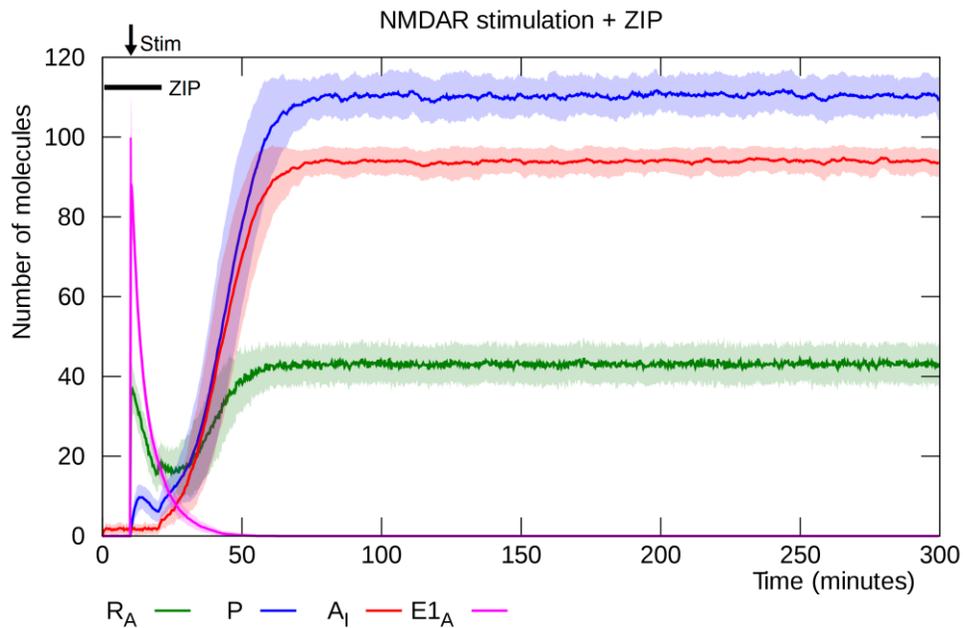

**Fig 6: ZIP immediately after stimulation does not prevent L-LTP induction.** In this simulation, ZIP inhibits PKMζ activity during the first 10 minutes after stimulation. L-LTP induction is delayed somewhat compared to Fig 3, but enough active PKMζ mRNA remains when the ZIP is removed to trigger a transition to the potentiation state. Solid lines represent mean molecule counts for 100 simulations. Lightly colored bands indicate standard deviation. $R_A$: active PKMζ mRNA, P: PKMζ, $A_I$: inserted GluA2-containing AMPARs, $E1_A$: activated E1 enzyme.

Presence of ZIP during the first ten minutes after stimulation does not prevent L-LTP induction [65]. The stimulation lifts the translational block and PKMζ production gets started. Even though PKMζ's enzymatic activity is inhibited, the mRNA stays activated long enough to ride out the ZIP activity. When the ZIP is washed out, PKMζ becomes active and drives the synapse into its potentiated state.

## PKMζ infusion induces L-LTP

L-LTP can be induced by diffusion of PKMζ into a neuron [39,41]. We simulate infusion by rapidly increasing the number of PKMζ molecules in the synaptic compartment to 100. This causes the model to settle into its potentiated state, (Fig 7).

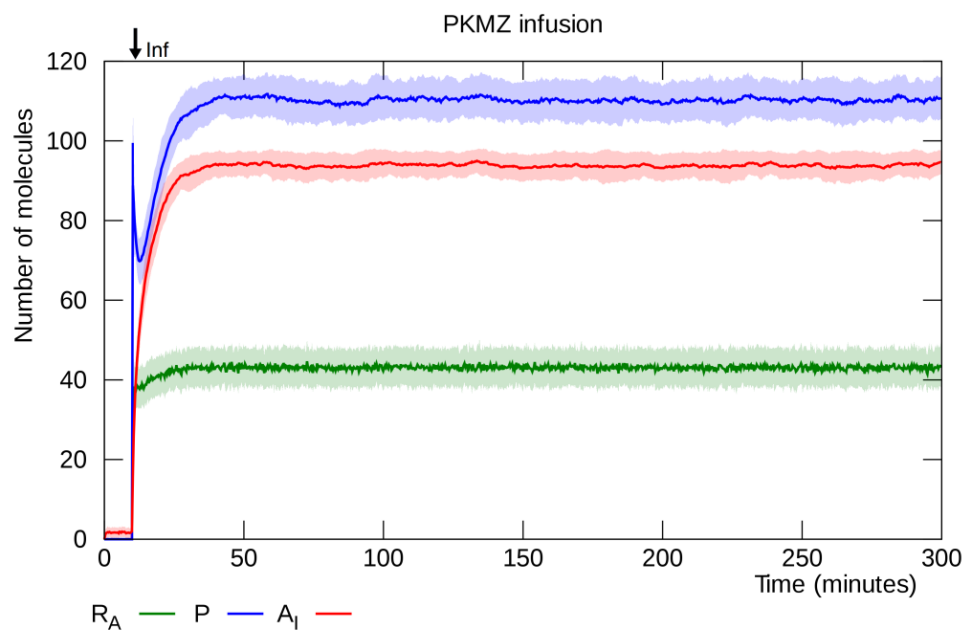

**Fig 7: PKMζ infusion induces L-LTP.** Infusion is simulated by stepping the PKMζ molecule count to 100 at "Inf". The PKMζ lifts the translational inhibition of PKMζ mRNA, synthesis starts and the synapse switches to its potentiated state. Solid lines represent mean molecule counts for 100 simulations. Lightly colored bands indicate standard deviation. $R_A$: active PKMζ mRNA, P: PKMζ, $A_I$: inserted GluA2-containing AMPARs.

## PSI blocks PKMζ-infusion-induced potentiation

The same level of PKMζ infusion that induces L-LTP in the previous experiment (100 molecules) fails to do so in the presence of PSI (Fig 8). Although the PKMζ infusion initially causes a temporary increase in the number of inserted GluA2-containing

AMPARs, the PSI prevents replenishment to compensate for PKMζ degradation and dissipation and the model returns to its unpotentiated state. This result, though plausible, has not been demonstrated in a published experiment. It thus constitutes a prediction of the model.

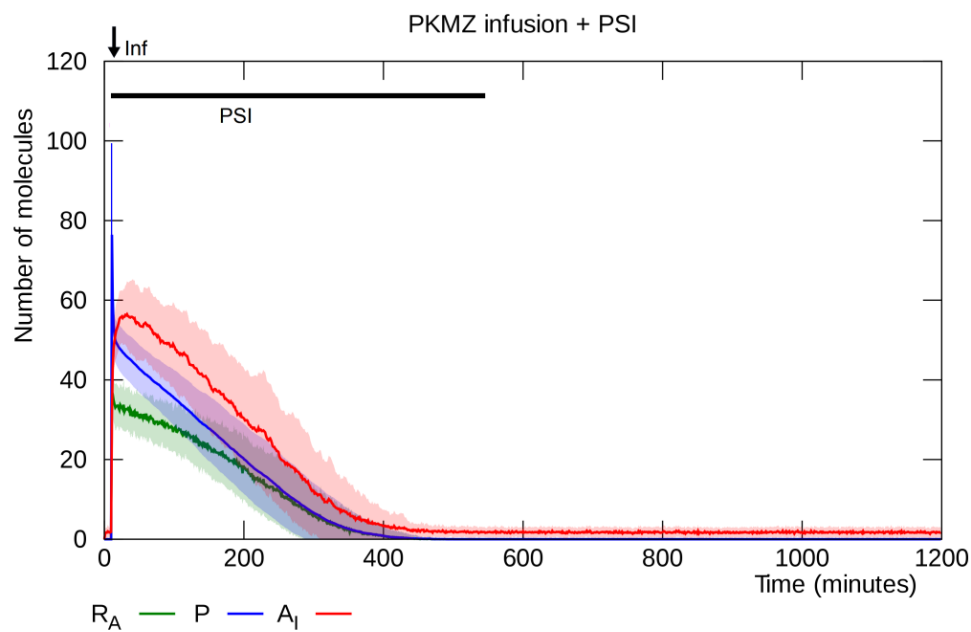

**Fig 8: PSI blocks L-LTP induction by PKMζ infusion.** As in Fig 7, infusion of PKMζ is simulated at "Inf". PKMζ triggers activation of PKMζ mRNA as well as an increase of inserted GluA2-containing AMPARs, but in the absence of PKMζ synthesis (blocked by PSI), the PKMζ level declines and the synapse settles back into its unpotentiated state. Solid lines represent mean molecule counts for 100 simulations. Lightly colored bands indicate standard deviation. $R_A$: active PKMζ mRNA, P: PKMζ, $A_I$: inserted GluA2-containing AMPARs.

## PSI does not disrupt established L-LTP

Fonseca et al. [12] demonstrated that suppressing protein synthesis for 100 minutes by bath application of anisomycin did not disrupt established L-LTP. Fig 9 shows the results of simulating this experiment in our model. The interruption of protein synthesis causes the number of PKMζ molecules to drop, which in turn leads to a transient decline in the number of inserted GluA2-containing AMPARs, but the system recovers when the PSI is removed.

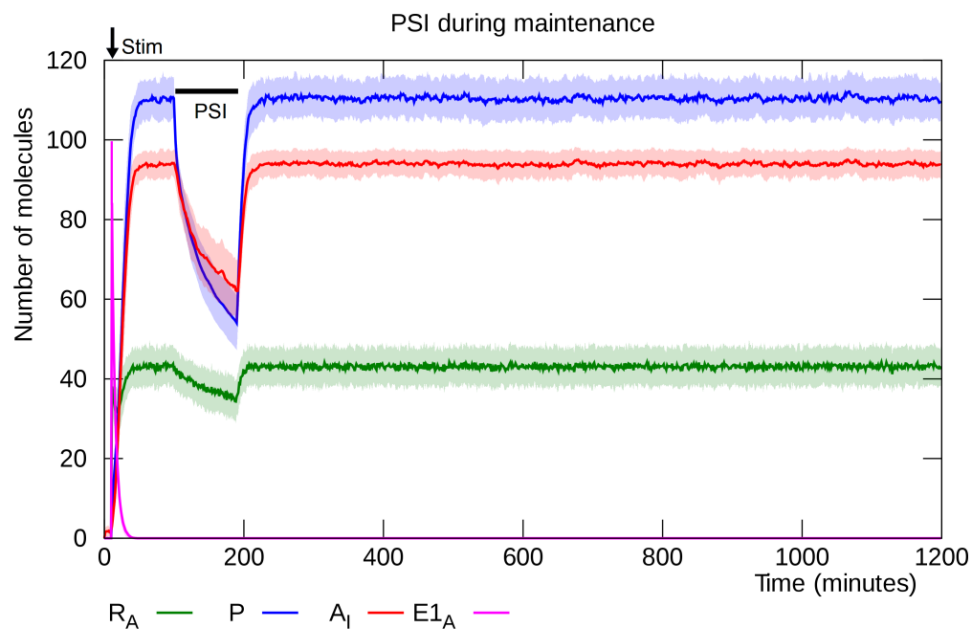

**Fig 9: PSI infusion during the maintenance phase does not disrupt established L-LTP.** L-LTP is induced by NMDAR stimulation at "Stim". Once L-LTP is established (100 minutes after induction), protein synthesis inhibition is applied for 100 minutes. The interruption of kinase synthesis causes a decline in the levels of PKMζ and inserted GluA2-containing AMPARs, but the synapse recovers when the PSI is removed. Solid lines represent mean molecule counts for 100 simulations. Lightly colored bands

indicate standard deviation. RA: active PKMζ mRNA, P: PKMζ, $A_I$: inserted GluA2-containing AMPARs, $E1_A$: activated E1 enzyme.

If the model is correct, then the transient decrease in the number of GluA2-containing AMPARs may be detectable as a reduced EPSP current after PSI application. However, it is possible that the temporary removal of GluA2-containing AMPARs is compensated for by insertion of GluA2-lacking AMPARs, similarly to what has been shown to happen during retrieval-induced destabilization [45], in which case the synaptic strength would be maintained. If this is the case, then it may instead be possible to detect a transient increase in rectification index, because GluA2-lacking AMPARs, but not GluA2-contaning ones, are characterized by a slight inward rectification [20,45]. Our model thus predicts that one or the other of these two effects (EPSP reduction or rectification) should be detectable after PSI application during L-LTP maintenance.

## Reactivation destabilizes, but does not disrupt, L-LTP

The effect of memory reactivation is simulated as a brief spike in the amount of active E2 enzyme (Fig 10). This results in rapid endocytosis of the inserted GluA2-containing AMPARs [20,21] and release of the bound PKMζ molecules which then start to dissipate. However, due to continued synthesis, the PKMζ level is kept from dropping below threshold and the model settles back into the potentiated steady state [18,79].

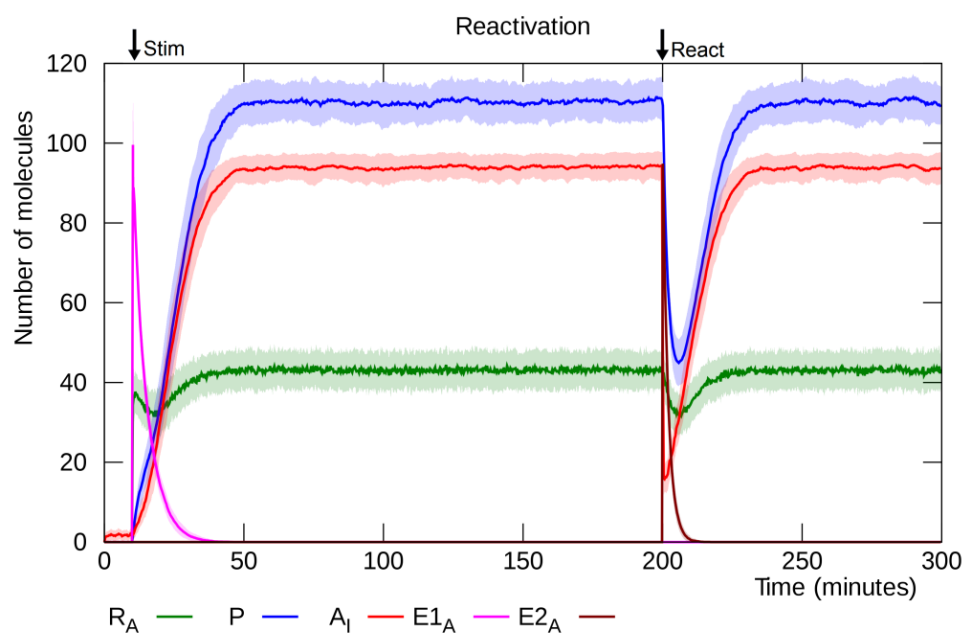

**Fig 10: Reactivation.** NMDAR stimulation is simulated by a pulse of active E1 enzyme at "Stim", and reactivation by a pulse of active E2 enzyme at "React". E2$_A$ causes rapid endocytosis of GluA2-containing AMPARs, which in turn leads to PKMζ depletion. PKMζ mRNA only declines slowly, however, and the synapse returns to its potentiated state when the E2 enzyme deactivates. Solid lines represent mean molecule counts for 100 simulations. Lightly colored bands indicate standard deviation. R$_A$: active PKMζ mRNA, P: PKMζ, A$_I$: inserted GluA2-containing AMPARs, E1$_A$: activated E1 enzyme, E2$_A$: activated E2 enzyme.

Although the population of inserted GluA2-containing AMPARs is almost completely depleted after reactivation, the levels of PKMζ and active PKMζ mRNA stay well above their depotentiation thresholds and the model reliably recovers from post-reactivation instability (reconsolidation), unless challenged by simulated pharmacological interventions (see below). As mentioned earlier, Hong et al. demonstrated this abrupt decrease of inserted GluA2-containing AMPARs after memory retrieval, as well as a corresponding transient increase of GluA2-lacking AMPARs, which maintained the synaptic strength during the labile period [20].

## Reactivation followed by PSI disrupts L-LTP

Simulation of PSI infusion simultaneously with reactivation, or shortly thereafter, causes disruption of L-LTP (Fig 11).

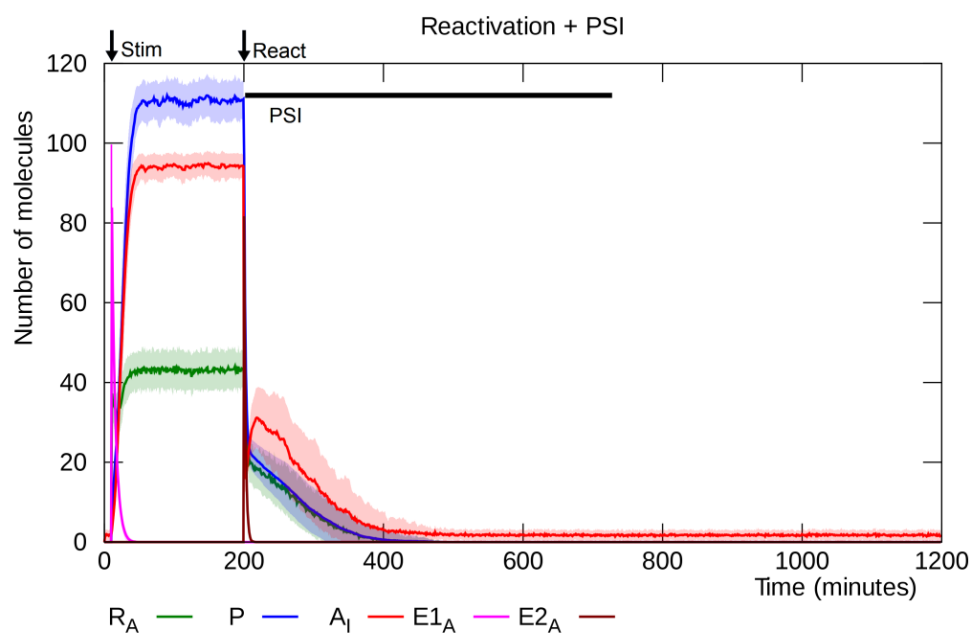

**Fig 11: Reactivation with simultaneous PSI infusion.** As in Fig 10, reactivation is simulated as a pulse of active E2 enzyme at "React", but here the presence of PSI prevents recovery and L-LTP is disrupted. Solid lines represent mean molecule counts for 100 simulations. Lightly colored bands indicate standard deviation. $R_A$: active PKMζ mRNA, P: PKMζ, $A_I$: inserted GluA2-containing AMPARs, $E1_A$: activated E1 enzyme, $E2_A$: activated E2 enzyme.

In the absence of new protein synthesis, the PKMζ level drops below threshold and the model settles into its unpotentiated state [18,79]. By varying the delay between reactivation and PSI infusion, we can establish the model's reconsolidation window, the time interval after reactivation during which L-LTP is vulnerable to PSI. As shown in Fig 12, if PSI infusion is applied 15 minutes or less after reactivation, then the model reliably switches to its lower (unpotentiated) steady state with few inserted GluA2-containing AMPARs, but with a delay of 30 minutes or more, L-LTP disruption does not result: the model remains in its potentiated state where the number of inserted GluA2-containing AMPARs fluctuates in the 60-100 range. The model's reconsolidation window is thus in the range 20 to 30 minutes, consistent with empirical results [18,84].

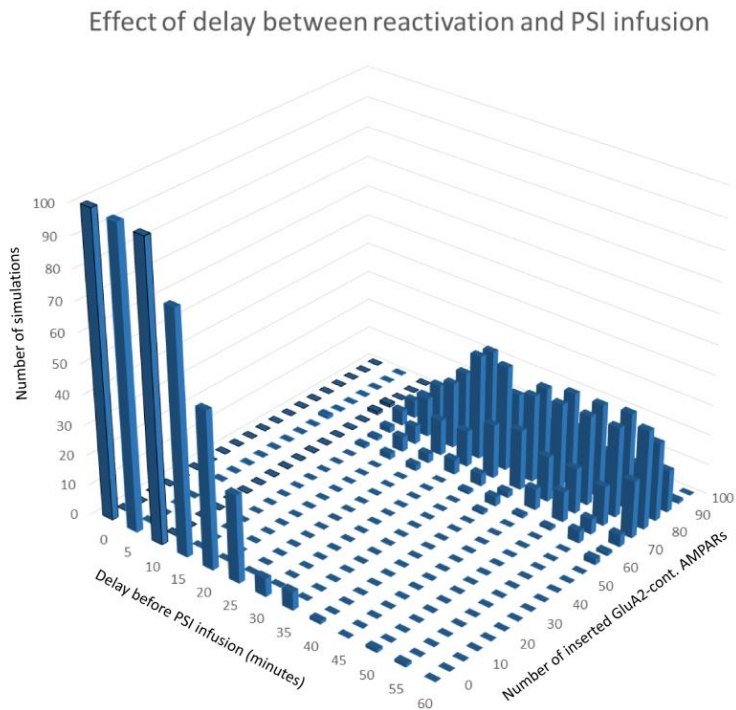

**Fig 12: Reconsolidation window.** Results of simulated reactivation followed by PSI infusion. The delay between reactivation and PSI infusion is varied from 0 to 60 minutes in 5-minute steps. One hundred simulations were run with each value for the delay. The number of inserted GluA2-containing AMPARs was recorded twenty hours after stimulation. For each value of the delay, the heights of the columns indicate the number of simulations that terminated with the corresponding numbers of inserted AMPARs.

## GluA2$_{3Y}$ blocks post-reactivation PSI-infusion from causing depotentiation

When the GluA2$_{3Y}$ peptide is infused together with PSI after reactivation, it prevents the disruption of L-LTP that PSI otherwise causes [57,63].

As before, reactivation triggers activation of the E2 enzyme, but here the GluA2$_{3Y}$ peptide blocks its endocytotic effect. As a result, the GluA2-containing AMPARs remain inserted and although the PSI stops synthesis of new PKMζ, the existing population of

PKMζ molecules, bound to the inserted GluA2-containing AMPARs, declines at a slow

enough rate to maintain the synapse in its potentiated state while the PSI wears off

(Fig 13).

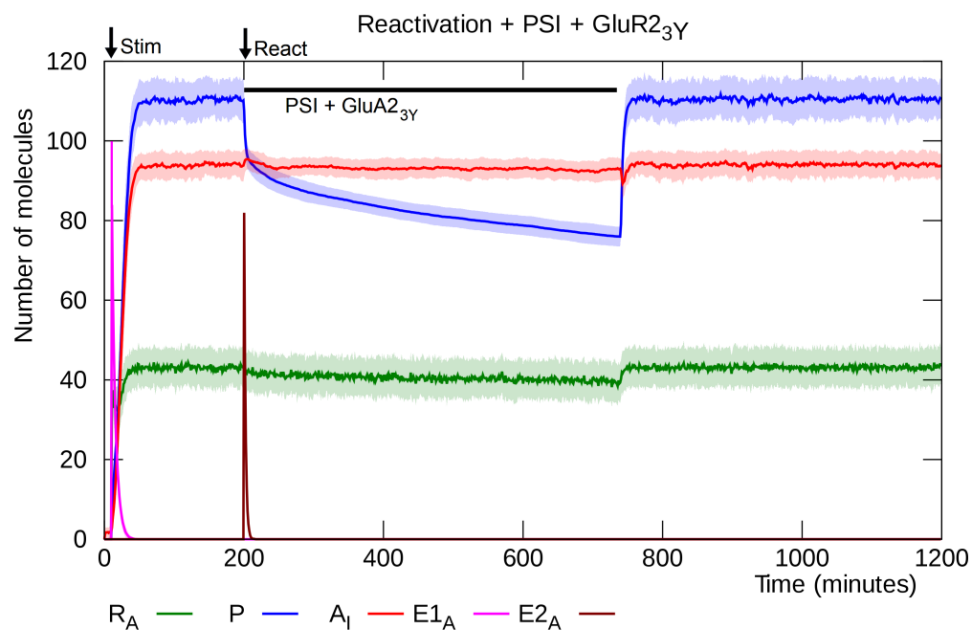

**Fig 13: Infusion of PSI and GluA2₃Y immediately after reactivation.** In this simulation

the endocytotic effect of the reactivation-triggered pulse of active E2 enzyme is blocked

by GLUA2₃Y. As a result, the GluA2-containing AMPARs remain inserted and continue

to sequester PKMζ molecules. The post-reactivation application of PSI still causes a

decline in the level of PKMζ, but because of the low dissipation/degradation rate, the

PKMζ level remains high enough that the L-LTP survives until the PSI wears off. Solid

lines represent mean molecule counts for 100 simulations. Lightly colored bands

indicate standard deviation. $R_A$: active PKMζ mRNA, P: PKMζ, $A_I$: inserted GluA2-

containing AMPARs, $E1_A$: activated E1 enzyme, $E2_A$: activated E2 enzyme.

## ZIP infusion disrupts established L-LTP

Infusion of ZIP during L-LTP maintenance causes rapid depotentiation [39–41]. ZIP inhibits PKMζ enzymatic activity, including both the catalysis of its own synthesis and the maintenance of an increased level of inserted GluA2-containing AMPARs in the PSD. The result is rapid removal of GluA2-containing AMPARs and depletion of PKMζ, and the synapse quickly settles into its unpotentiated state (Fig 14). The minimum duration of ZIP application needed to reliably disrupt L-LTP in the model is around 30 minutes.

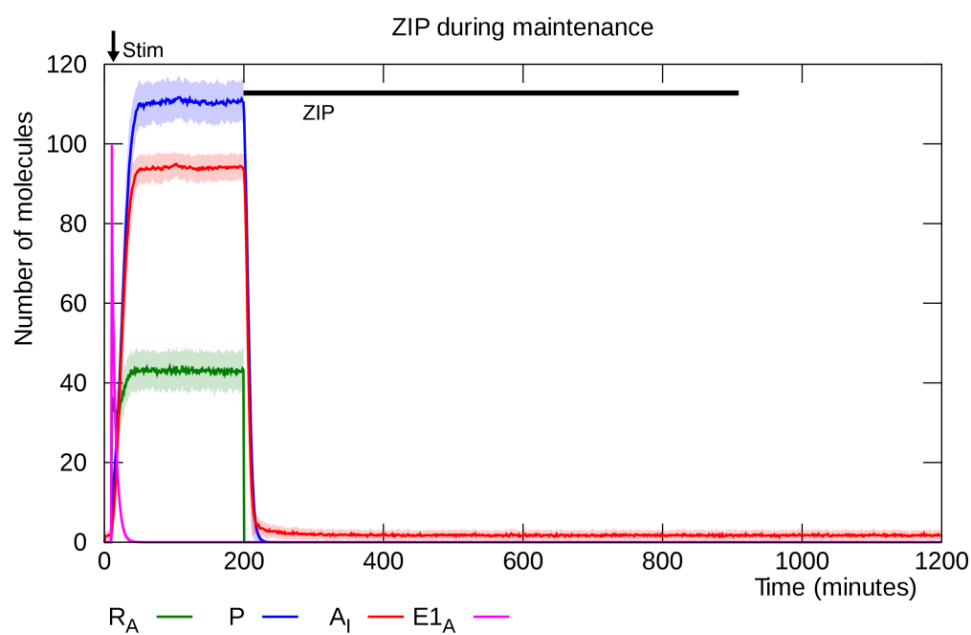

**Fig 14: ZIP infusion during L-LTP maintenance.** Application of ZIP inhibits PKMζ's enzymatic activity, leading to rapid depotentiation. Solid lines represent mean molecule counts for 100 simulations. Lightly colored bands indicate standard deviation. $R_A$: active PKMζ mRNA, P: PKMζ, $A_I$: inserted GluA2-containing AMPARs, $E1_A$: activated E1 enzyme.

## GluA2_{3Y} blocks depotentiation by ZIP infusion

When the GluA2_{3Y} peptide is infused together with ZIP during L-LTP maintenance, the disruptive effect of ZIP is blocked [33].

As before, ZIP inhibits PKMζ's catalysis of its own synthesis as well as its facilitation of AMPAR trafficking into the PSD and its blocking effect on BRAG2-induced endocytosis of GluA2-containing AMPAR. But in this case, even though BRAG2 remains active, the presence of GluA2_{3Y} prevents it from inducing endocytosis of the inserted GluA2-containing AMPARs. As a result, the GluA2-containing AMPARs remain in the PSD and continue to maintain the PKMζ molecules at the synapse. The number of PKMζ molecules declines only slowly and the potentiation is able to survive through the 12-hour period of ZIP activity (Fig 15).

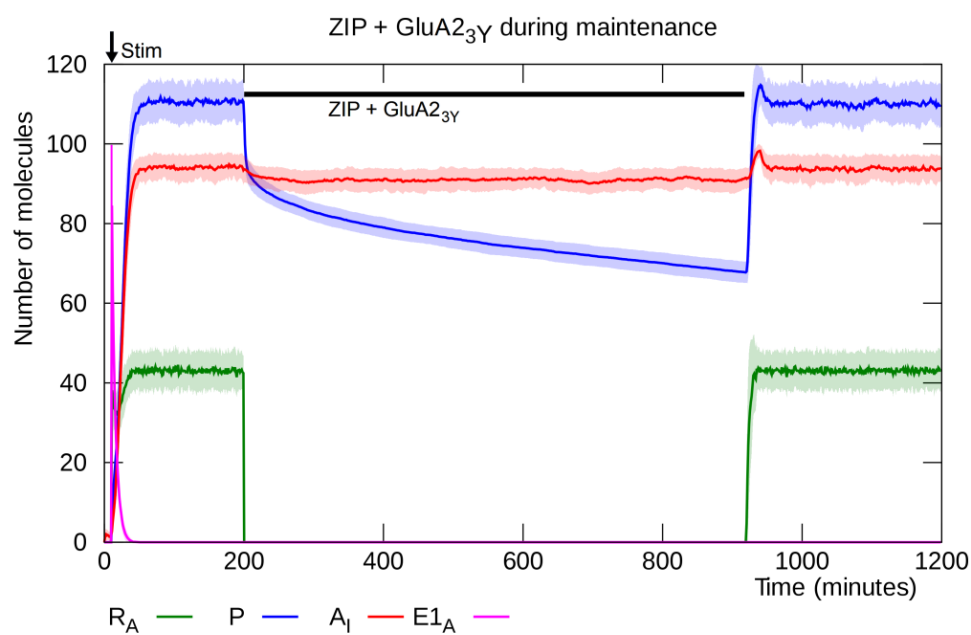

**Fig 15: Infusion of ZIP and GluA2_{3Y} during L-LTP maintenance.** ZIP blocks PKMζ's enzymatic activity: PKMζ mRNA returns to its untranslatable state, and BRAG2

becomes active. However, GluA2$_{3Y}$ prevents BRAG2 from inducing GluA2-containing

AMPAR endocytosis, the PKMζ molecules remain attached to the inserted AMPARs,

and the catastrophic disruption of L-LTP seen in Fig 14 is averted. Solid lines represent mean molecule counts for 100 simulations. Lightly colored bands indicate standard deviation. $R_A$: active PKMζ mRNA, P: PKMζ, $A_I$: inserted GluA2-containing AMPARs,

$E1_A$: activated E1 enzyme.

# Discussion

The model presented here is able to explain a range of results relating to the role of

PKMζ in late-phase long-term synaptic potentiation, including L-LTP induction by

NMDAR stimulation or by PKMζ infusion and the findings that whereas PSI, but not ZIP,

can block induction of L-LTP, the reverse is true for disruption of established L-LTP. In

addition, it accounts for cellular reconsolidation, reconsolidation blockade by PSI

infusion and prevention of ZIP- or PSI-induced depotentiation by infusion of the GluA2$_{3Y}$

peptide. While subsets of these results have been covered by earlier models [85–89],

ours is the first to account for all of them. A further distinguishing feature of our model is

that it demonstrates that a wide range of empirical findings described in the LTP

literature can be accounted for by simple molecular reactions whose rates are governed only by the law of mass action, i.e. without postulating cooperative binding or other non-linear dependencies on reactant concentrations.

Our model demonstrates that a bistable mechanism for synaptic potentiation can arise from the interaction of two coupled feedback loops, neither of which needs itself be bistable. One of these, the mutual reinforcement between PKMζ and PKMζ mRNA, has been featured in previously published models of L-LTP maintenance [85–88]. The second positive feedback relationship in our model is between PKMζ and inserted GluA2-containing AMPARs, which mutually maintain each other by inhibiting each other's removal from the synapse [42]. The ability of inserted AMPARs to sequester PKMζ molecules at the synapse allows the model to account for findings involving the inhibition of regulated endocytosis of GluA2-containing AMPAR [33,57].

Our model exhibits robust bistability; when left to run for a full year of simulated time in either the potentiated or depotentiated state, no spontaneous transitions between the steady-states were observed. The source of this bistability can be understood by considering the interaction between the two feedback loops. The PKMζ-mRNA interaction is a positive feedback loop: A greater number of PKMζ molecules will keep more mRNA molecules in an unrepressed state and more unrepressed mRNA results in a higher rate of PKMζ synthesis. This subsystem has two steady states: a lower steady state with zero PKMζ molecules and zero unrepressed mRNA molecules, and a higher state at a level that depends on the reaction rates, in particular PKMζ's dissipation rate, because at equilibrium the synthesis and dissipation rates are equal. The lower steady state is unstable; the introduction of just a few PKMζ molecules can cause a switch to the upper state. The PKMζ – mRNA feedback loop thus has only a single stable steady state which depends on the PKMζ dissipation rate, as illustrated in Fig 16.

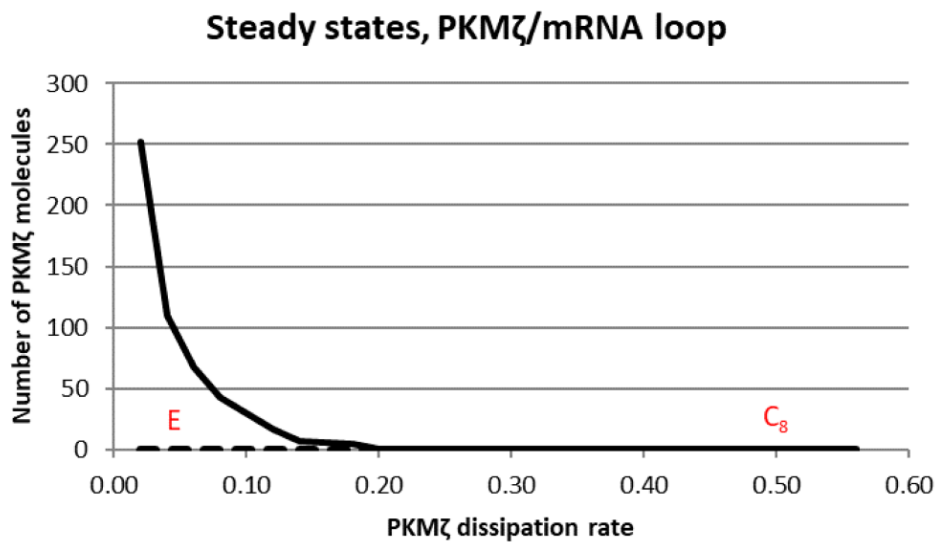

**Fig 16: PKMζ level at steady state as a function of dissipation rate.** The solid line represents a stable steady state, and the dashed line an unstable steady state. The x-axis represents the reaction constant for PKMζ dissipation/degradation. "$C_8$" indicates the value used for the reaction constant of reaction 8, dissipation/degradation of unbound PKMζ. "E" indicates the effective dissipation rate in the potentiated state, when a large proportion of the PKMζ molecules are bound to inserted GluA2-containing AMPARs.

Bistability arises because of the influence of the second feedback loop, the interaction between PKMζ and GluA2-containing AMPARs. In the unpotentiated state, the PKMζ dissipation/degradation rate is controlled by the reaction constant $c_8$, which has a value of 0.5. As seen in Fig 16, the steady state at this rate has zero PKMζ molecules. In the potentiated state, an increased number of GluA2-containing AMPARs in the PSD bind

PKMζ molecules; this results in a reduction of the effective PKMζ dissipation/degradation rate to a value where the steady state has ca 100 PKMζ molecules (indicated by 'E' in Fig 16).

## Comparison with previous computational models of PKMζ regulation

Clopath et al. [89] describe a mathematical model of synaptic tagging and capture (STC) [90], wherein mechanisms of tag-setting and triggering of protein synthesis interact with a bistable process that maintains potentiation. Although the authors suggest that one of the model's parameters may represent the level of PKMζ activity, the mechanisms of the process are unspecified, and the model therefore cannot account for the results targeted by our model: the effects of PSI, ZIP and GluA2$_{3Y}$ in the contexts of L-LTP induction and maintenance, or of memory reactivation.

A simple model by Ogasawara and Kawato [86] simulates L-LTP induction and maintenance as well as reconsolidation based on the interactions of only three molecules: PKMζ, PKMζ mRNA and F-Actin. It is, however, not able to account for most of the results addressed in this paper.

A paper by Zhang et al. [88] features a dual-loop model of LTP that exhibits windows of susceptibility to PSI after induction and reactivation as well as vulnerability to a kinase inhibitor in the maintenance phase. The relationship between the kinase and AMPA receptors is not modeled, and thus the ability of an endocytosis blocker like GluA2$_{3Y}$ to rescue L-LTP is not accounted for. Also, the kinase modeled in [88] is unnamed but characterized by auto-activation rather than persistent activity, and should therefore probably not be interpreted as PKMζ.

Smolen et al. [87] model synaptic tagging and capture, including "cross-tagging" between LTP and LTD. As in our model, synaptic stability is based on PKMζ's ability to catalyze its own synthesis. Unlike our model, [87] does not account for the effects of protein synthesis inhibition, kinase inhibition, reactivation or the ability of endocytosis blocking to rescue L-LTP.

A paper by Jalil et al. [85] models PKMζ regulation at the synapse, with a focus on compensatory interactions between PKMζ and a second atypical PKC isoform, PKC$_{\iota/\lambda}$. Bistability is achieved by combining the PKMζ auto-catalytic synthesis feedback loop with auto-phosphorylation. The model predicts the differential effects of ZIP and PSI at L-LTP induction and maintenance, but does not account for L-LTP rescue by AMPAR endocytosis blocking, nor for reconsolidation.

## Limitations

Our model represents a subset of the mechanisms believed to be involved in LTP induction and maintenance [3,91]. Some processes not included in our model are:

- the induction and stabilization of early LTP, which likely involves GluA2-lacking AMPARs [45], the MAPK/ERK signaling pathway and the proteins PKA, CaMKII [91] and PKCλ [65,92]

- a later phase of L-LTP, sometimes called LTP3, which requires gene transcription as well as mRNA translation [93] and may involve a "tagging and capture" mechanism for selectively targeting gene products to potentiated synapses [40,90].

- polymerization/depolymerization of actin and restructuring of the cytoskeleton [94,95]

The processes that we have modeled thus form a subset of a more complex machinery. Nevertheless, it is interesting to note that this relatively simple model is able to account for many of the empirical findings regarding the role of PKMζ in L-LTP induction and maintenance, and to exhibit the degree of stability required for a neural mechanism to support long-lasting memories.

## Acknowledgments

We thank Marcel Montrey for helpful comments and suggestions.

# Supporting Information

**S1 Text. Estimating the number of PKMζ molecules in a spine head.**